\documentclass[12pts,manuscript,a4paper]{aastex}
\usepackage{graphicx}
\usepackage{graphics}
\usepackage[english]{babel}
\usepackage[tight,scriptsize]{subfigure}
\shorttitle{UV Radiation Constraints in Habitable Zones}
\shortauthors{Buccino et al.}

\begin{document}

\renewcommand{\baselinestretch}{1}
\date{}
\titlepage

\title{Ultraviolet Radiation Constraints around the Circumstellar Habitable  Zones}

\author{\textbf{Andrea P. Buccino}}
\affil{\small{ Instituto de Astronom\'\i a y F\'\i sica del Espacio (CONICET),
    C.C. 67 Sucursal 28, C1428EHA-Buenos Aires Argentina}}
\vspace{-1cm}
\email{abuccino@iafe.uba.ar}

\author{\textbf{Guillermo A. Lemarchand}}
\affil{\small{Facultad de Ciencias Exactas y Naturales, Universidad
    de Buenos  Aires,
    C.C. 8 - Sucursal 25, C1425FFJ Buenos Aires Argentina and Instituto
    Argentino de Radioastronom\'\i a (CONICET)}}

\author{\textbf{Pablo J. D. Mauas}}
\affil{\small{Instituto de Astronom\'\i a y F\'\i sica del Espacio (CONICET),
    C.C. 67 Sucursal 28, C1428EHA-Buenos Aires Argentina}}
\received{Decembre, 9th 2005 by Icarus}

\accepted{March, 28th 2006 by Icarus.}

\newpage
\begin{abstract}

 Ultraviolet radiation is known to inhibit photosynthesis,
induce DNA destruction and cause damage to a wide variety of
proteins and lipids. In particular, UV radiation between 200-300 nm
 becomes energetically very
damaging to most of the terrestrial biological systems.
 On the other hand, UV  radiation is usually considered one of the most
 important  energy source on the
primitive  Earth for the synthesis of many biochemical compounds and,
therefore,  essential for several biogenesis
processes.  In this work, we use these properties of the UV radiation to
define the boundaries of an
ultraviolet habitable zone. 
We also analyze the evolution of the
UV habitable zone during the main sequence stage of the
star.

We apply these criteria to study the  UV habitable zone for those
extrasolar planetary systems that were observed by the
\emph{International Ultraviolet Explorer (IUE)}.  We analyze the
possibility  that
extrasolar planets and moons could be suitable for life, according to
the UV  constrains presented in this work and other
accepted criteria of habitability (liquid water, orbital stability,
etc.).

\end{abstract}
\renewcommand{\baselinestretch}{1.}
\section{Introduction}\label{intro}
The so-called ``Principle of Mediocrity" proposes that our planetary
system, life on Earth and our technological civilization are about average and
that  life
and intelligence will develop by the same rules of natural selection wherever
the proper conditions and the needed time are given (von Hoerner
1961, 1973).
In other words, the conditions that give place to the origin
and  evolution
of life on Earth are average, in
comparison to other worlds in the universe.

This hypothesis is
in the  ``hard core'' (Lakatos 1974) of all the research programs that
search for life
in the universe, which during the last fifty years were known within the
scientific community as exobiology, bioastronomy, astrobiology, CETI, SETI,
etc. 

Using the ``Principle of Mediocrity'', we speculated about
the  existence of
Earth-like planets, which must have liquid water on its surface,
comparable surface inventories of CO$_{2}$, H$_{2}$O, N$_{2}$ and other
biogenic elements, an early history allowing chemical evolution that leads to
life, and subsequent climatic stability
 for at least 4.5 billion years
(Lineweaver 2001; Owen 2000).  We also speculated about the
 possible  universal
mechanisms for the origin of life, about universal mechanisms of
Darwinian natural selection and for the appearance of intelligence and
technological civilizations, and about how to detect primitive life
and advanced technological civilizations beyond our home planet (Shklovskii and
Sagan 1966; Lemarchand 1992).

Even though there are plans to do so (Fridlund 2000, 2002; Mennesson
\emph{et al.} 2005),
State-of-the-Art technologies do not allow us  yet to detect Earth-like
extrasolar planets, and  even less to analyze the composition of their
atmospheres  in
detail. Recently, Vidal-Madjar \emph{et al.} (2004) observed
transits of the planet orbiting the star HD209458, and for the first
time detected  the existence of oxygen and carbon in an extrasolar planet's
atmosphere, which is a step toward the confirmation that our home planet may be
``average''.

Stars which are similar to the Sun in mass and evolutionary state,
i. e. that have broadly a similar physical structure, are usually
called \emph{solar-like} stars. In practical terms, Soderblom and King
(1998)  defined \emph{solar-like} as main sequence stars of
spectral class F8V to K2V (or 0.50 to 1.00 in B$-$V). Several
extrasolar  planets
have been found around this type of stars (Schneider 2005), and it
is important
to study whether they are suitable for life.

The original concept of Habitable Zone (HZ)  was introduced  by Huang
(1959, 1960) and  was later extended for
the habitability by complex and intelligent life  in a seminal study
by  Dole (1964). Later, Hart (1978, 1979) calculated the
hypothetical  evolution of the
terrestrial atmosphere on geological time-scales for different orbital
 radius.  He found
that the HZ between runaway greenhouse and runaway glaciation processes is
amazingly narrow for G2 stars like the Sun. 

These early
calculations omitted the negative feedback between
atmospheric  CO$_2$
partial pressure and mean global surface temperature via the carbonate-silicate
cycle discovered by Walker \emph{et al.} (1981).
Taking into account this feedback, Kasting \emph{et
    al.} (1988)  found the  interesting result of an almost constant
  inner boundary and a remarkable extension of the outer boundary.

In a comprehensive study, using a 1-D
radiative-convective climate model,
Kasting \emph{ et al.} (1993) presented a refined calculation of the
HZ for  the
solar system and for other main sequence stars. They studied the effects of
the loss of water via photolysis and of hydrogen escape, and the
formation of CO$_2$ clouds, which cools the planet's surface by increasing its
albedo and by lowering the convective lapse rate. They also showed how
climatic stability is ensured by a feedback mechanism in which
atmospheric  CO$_2$
vary inversely with planetary surface temperature. For the internal HZ
boundary,
they give three different estimations, ranging from 0.75 to 0.95
AU. The first case  assumes the loss of planetary
water by a moist greenhouse (Kasting 1988), the second one assumes loss of
planetary water by a runaway greenhouse and the third one is based on the
observation that there has been no liquid water on Venus' surface at
least during the
last Gyr. The external HZ boundary was also estimated in three different ways:
the first criterion considers the characteristics of the evolution of the early
martian atmosphere (Kasting 1991), the second one assumes a maximum possible
CO$_2$ greenhouse heating, and the third one is related to the first
condensation limit of CO$_2$ clouds that increase the planetary albedo. These
criteria lead to an external boundary in a range from 1.37 to 1.77 AU.

Other more general criteria to understand the optimal position of the HZ 
was proposed by Franck \emph{et al.} (2000a,
2000b).
Based on the global carbon
cycle as mediated by life and driven by increasing solar luminosity and plate
tectonics, they estimated the biogeophysical domain supporting a
photosynthesis-based ecosphere during planetary history and
future. They corroborated the  mediocrity assumption, since they found
that the optimum position where the environment would
accomplish the maximum life span and biological
productivity\footnote{The  biological productivity is the amount of
  biomass  that is produced by photosynthesis per unit of time and per
  unit  of continental area.} is at 1.08 AU, very close to the radius
of Earth's orbit.  

The evolution in time of stellar luminosity is another important
factor to  determine
the location and the size of circumstellar habitable zones (Whitmire and
Reynolds 1996). Habitable zones tend to migrate outward with time because main
sequence stars become brighter as they age and convert hydrogen into helium and
heavier elements. The continuously habitable zone (CHZ) is the region
in space  where a planet remains habitable for some
specified long time period ($\tau_{hab}$).  A common choice for this time is
$\tau_{hab}$$\sim$ 4 Gyr, the  time that was needed on Earth for
intelligent  life to
emerge and evolve into a technological civilization. Henry
\emph{et al.} (1995) and Turnbull and Tarter (2003a) considered
$\tau_{hab}$$\sim$ 3 Gyr, while Schopf (1993) used  the
time required for microbial life to emerge $\tau_{M life} \sim$ 1 Gyr
in his   definition of the CHZ.

Recently, Turnbull and Tarter (2003a, 2003b) used the HZ
 definition, together with other variables like  X-ray
luminosity, Ca\,\scriptsize{II}\normalsize\- H and K activity,
rotation,  spectral
types, kinematics, metallicity and Str\"omgren photometry, to build a
 catalog of nearby stars suitable for the
development of life and technological civilizations (``HabCat'').

The ultraviolet radiation emitted by a  star can also be 
important  to determine the
suitability of
extrasolar planets for biological evolution and for the subsequent
adaptation of  life in exposed habitats (Cockell 1999). This factor
will also  determine the "average"
environmental conditions that we think will help extraterrestrial life
to  develop on an Earth-like planet.

Ultraviolet radiation has also played a key role in the evolution of the terrestrial atmosphere. Guinan
\emph{et al} (2003) showed that the strong FUV and UV emissions of the young Sun could have played a major
role in the early development and evolution of planetary atmospheres. 
X-ray and EUV emission fluxes of the
early Sun could have also produced significant heating of the
planetary exosphere and upper-atmosphere thus enhancing processes such
as thermal escape (e.g. Luhmann and Bauer 1992) and could also played a
role in the origin and development of life on Earth and possibly on
Mars.  The high levels of FUV radiation of the Sun could strongly
influence  the abundance of NH$_3$,
CH$_4$  and O$_3$ in the prebiotic and
Archean terrestrial atmosphere some 2-4 Gyr ago. Ozone is an efficient
screening mechanism for the
enhanced UV radiation, thus protecting the emerging and evolution of
life on the surface planet. 

In previous works (Buccino \emph{et al.} 2002, 2004), we have shown
that UVB and UVC radiation may have its own boundary  conditions to
fill the biological  requirements of an habitable zone.
This effect may be useful to define a new boundary condition for the
UV-HZ inner limit.

On the other hand, besides the fact that ultraviolet radiation plays an 
important role destroying life,
it was also one of the most important energy source on the
primitive  Earth for the synthesis of many biochemical compounds that
may  derivate from HCN (Toupance \emph{et al.} 1977) and
essential for some biogenesis
processes (Sagan 1973).  The need of a certain level of ultraviolet
radiation to start several biogenic
processes, naturally defines an outer border
for an  UV-HZ. 

In this work, we use these concepts in a new definition of an
ultraviolet habitable zone, and apply it to all nearby stars harbouring
exoplanets that have been observed in the UV by the \emph{International
  Ultraviolet  Explorer (IUE)}. 

In section
\ref{uvboundaries}, we introduce two constrains for the origin  and the
development of life. The inner limit is determined by the
levels of  UV damaging radiation tolerable by DNA and the outer
limit is characterized by the minimum UV radiation
needed in the biogenic process. 

In section \ref{uvradiation}, we apply these criteria to stars listed
  in the  \emph{Extrasolar
  Planets  Catalog} (Schneider 2005) and we analyze the possible
  exoplanets and  moons
that may be  suitable for life. Then, in section \ref{discussion}, we
analyze  the implication of the UV  radiation on the HZ defined in
  Kasting  \emph{et
  al.}  (1993) and we compare it with our
results.  We also discuss in that section the  factors that can
  alter the UV  constrains.

\section{Biological boundaries for UV habitable zones}\label{uvboundaries}

 The destructive effect of the UV radiation on biochemicals processes
 is  usually
considered through a biological action spectrum (BAS) $B(\lambda)$,
which  represents a
relative measure of damage as a function of wavelength. In this
work, we define $B(\lambda)$ as the probability of a photon of
energy $\frac{hc}{\lambda}$ to dissociate free DNA.

The first approximation of the action spectrum was obtained by Setlow
and Doyle (1954, see also Setlow 1974).  Cockell (1998) gave an
estimation of the  BAS based on studies by Green and Miller (1975)  
and Lindberg and Horneck (1991). We have
semi-empirically adjusted the curve  $B(\lambda)$ obtained from these
previous  studies
 with the following expression:
\begin{equation}
log\,B(\lambda)\sim
\frac{6.113}{1+\exp(\frac{\lambda[nm]-310.9}{8.8})}\,-4.6\,.
\end{equation}
\normalsize

Lindberg and Horneck (1991) showed that at wavelengths shorter than 230
 nm, the action spectra and the absorption spectrum
 of DNA diverge. This can
 be explained by assuming that radiation of much less than
 230 nm is absorbed by
 external layers of the spore cell wall surrounding
 the core in which the DNA is
 situated. Under these conditions only a very small
 fraction of shorter
 wavelengths reaches the DNA material. 

A measure of the DNA-damage caused by UV photons radiated
 by a star  of age $t$
 that reach the top of the atmosphere of a
 planet at a distance $d$ (in AU)  can therefore be expressed as

\begin{equation}       
N^{*}_{DNA}(d)=\int_{200 nm}^{315 nm} 
B(\lambda)\frac{\lambda}{hc}\,\frac{F(\lambda,t)}{d^2}d\lambda\,,
\label{eq.dna}
\end{equation}
where $F(\lambda,t)$ represents the UV flux of the star at 1 AU, $h$
is  the Planck
constant and $c$ is the speed of light.


If $N^{*}_{ DNA}(d)$ is larger than a certain threshold, it would
be  difficult for life to develop
on the planet's surface because of the damaging UV  radiation.
 The maximum amount of DNA-damage
that a  planet could surmount can be
larger than the one received by early Earth, where life has developed.

On the other hand, there  might be several 
sources of natural attenuation of the UV radiation, in particular a  planetary
atmosphere. Each
exoplanet might have different conditions that determine the exact
amount of  the UV attenuation over
the planetary surface. We quantified this attenuation with a factor
$\alpha$,  which is the ratio
between  the radiation
received  on the planetary surface and the incident radiation on top of the
atmosphere.

Segura \emph{et al.} (2003) performed a coupled
radiative-convective/photochemical model for Earth-like planets
orbiting G2V, F2V and K2V stars. They studied the
levels of UV radiation received on the planetary surface
considering  different
 O$_2$ concentrations.  For each atmospheric composition,
   therefore, we estimated the $\alpha$-factor as the ratio between
   the UVB+UVC fluxes received on the planetary surface and those with zero
   O$_2$ concentration. For 
10$^{-5}$ times the present values of O$_2$ at Earth, which could have been the
conditions of the Archean Earth,
the attenuation is  $\alpha$=0.84 for K and G stars and $\alpha$=0.74
for F stars.

For these reasons, and considering the complexity and
self-regulating characteristics of
life-systems, we apply  the  "Principle
of  Mediocrity" and  assume that the maximum UV radiation
tolerable, before DNA damaging prevents the appearance of life, is
equal to twice the radiation that reached  the outer space over the
Earth's atmosphere 3.8 million
years ago. We remark that the atmospheric corrections are contemplated
within  our
mediocrity factor.

Therefore, we will define the inner limit of the UV habitable zone as
the one  given by 
\begin{equation}
\hspace{-3cm}
N^{*}_{DNA}(d) \le 2 \times N^{\odot}_{DNA}(1 \textrm{\footnotesize{AU}\normalsize})|_{t=t^{\odot}_{Arc}},\label{innlim} 
\end{equation}
where
$N^{\odot}_{DNA}(1 \textrm{\footnotesize{AU}\normalsize})|_{t=t^{\odot}_{Arc}}$ is 
 calculated with Eq. \ref{eq.dna}, using the flux received by the
 Archean Earth ($\sim$ 3.8 Gyr ago), which we assumed equal to 75\% 
 the present radiation on top of the atmosphere, as was done by
 Kasting (1988)  and Cockell (1998). Both authors followed the Solar Standard
 Model predictions. 

On the other hand, UV radiation is  thought 
to have also played a positive role in the origin of life.
Toupance \emph{et al.} (1977), using early models of
  terrestrial  archean
atmosphere, showed that UV radiation may have
largely  contributed to the
synthesis of HCN in CH$_4$ - NH$_3$ and consequently to the
synthesis of many biochemical compounds that derive from
HCN.  
In the early Earth, methane-ammonia atmospheres are no longer
  considered realistic. Ammonia, in particular, photolyzes rapidly and
  is unlikely to have been present in the early Earth (Pavlov \emph{et al.}
  2001).

Molecular Hydrogen can also be generated from near UV irradiation of
aqueous ferrous hydroxide at  pH 6-8 and the presence of banded iron
formations in Archean rocks has been cited as evidence that Fe
photo-oxidation occurred on the early Earth (Braterman and Cairns-Smith 1987).

 Chyba and
McDonald (1995) described a possible non-atmospheric mechanism by which
ultraviolet photolysis could have acted as an energy source for prebiotic
organic chemistry.
Chyba and
Sagan (1997) and Ehrenfreund \emph{et
al.} (2002) proposed that the greater rate of production of organics
on  Earth, 4 Gyr ago, 
was due to UV photolysis.

Recently, Tian \emph{et al.} (2005) proposed a new model for
  the early terrestrial atmosphere that is CO$_2$-rich and not
  CH$_4$-rich, but contains as much as  30\% H$_2$. In their model,
  the production of organic  molecules through electrical discharges
  or ultraviolet light becomes  increasingly important relative to the
  delivery  of organic
molecules by  meteorites or comets. The organic soup in the oceans and ponds on
early Earth would have been more favorable places for the origin of
life
 than have been previously thought and, therefore,
UV light would have been more important
as an
 energy source for the formation of complex organic molecules.

Mulkidjanian \emph{et al.} (2003) presented another evidence of the importance
of UV radiation in biogenesis processes. They
simulated the formation of first oligonucleotides under continuous UV
illumination and  confirmed that UV irradiation could have
worked  as
a selective factor, leading to a relative enrichment of the system in longer
sugar-phosphate polymers carrying nitrogenous bases as UV-protectors.

Consequently, we assume that there is a  minimum number of UV
photons needed as an  energy source
for the  chemical
synthesis of complex molecules. The {\it total} number of UV
photons  received by a planet at distance $d$ can be computed as
\begin{equation}       
N^{*}_{UV}(d)=\int_{200 nm}^{315 nm}
\frac{\lambda}{hc}\,\frac{F(\lambda,t)} {d^2}d\lambda\,.
\label{eq.uv}
\end{equation}

Based on the "Principle of Mediocrity," we request that
the  planet receives
at least half the total number of UV photons received by the Archean Earth,
$N^{\odot}_{UV}(1\textrm{\footnotesize{AU}\normalsize})|_{t=t^{\odot}_{Arc}}$,
and we
set the  outer
limit of the UV habitable zone accordingly:

\begin{equation}
\hspace{-3cm}
N^{*}_{UV}(d) \ge 0.5 \times N^{\odot}_{UV}(1 \textrm{\footnotesize{AU}\normalsize})|_{t=t^{\odot}_{Arc}}.\label{outlim} 
\end{equation}

The mediocrity factor makes the UV habitable
zone  from 0.71 to 1.90 AU around
the Sun, wider 
than the traditional one proposed by Kasting \emph{et al.} (1993),
where the  limits
in the intermediate case, for the solar system, go from 0.84 to 1.67 AU in the solar system.

 
For life systems to evolve, the planet should be in the habitable zone
continuously during a certain time. 
To analyze the evolution of the habitable zones, we first assume that
 the  luminosities of  F, G and K main sequence stars
 evolve  in a similar way and that the UV radiation follows the same pattern.

According to Dorren and Guinan (1994) and Ribas \emph{et al.} (2005),
the  early Sun emitted a proportionally
higher flux at UV than at visible wavelengths. In particular, the UV continuum
flux density of the early Sun exceeded the present  one
by a factor of 2 to 10 shortward of 200 nm. However, at longer UV
wavelengths, where the
photospheric emission dominates over that of the chromosphere,
the UV flux of the young solar types would represent only 10-30\%
of that of the present Sun   (Ribas
\emph{et al.} 2005). Taking into
 account the biological
 influence of UV radiation, in this work we focused
 our study only in those
 wavelengths between 200-315 nm, where the evolution of UV
emission should scale with the bolometric solar luminosity.

For the evolution of the solar  luminosity in the past Gough
(1981) obtain the following  relation: 

\begin{equation}
L_{\odot}^{past}(t)=[1+\frac{2}{5}(1-\frac{t}{t_{\odot}})]^{-1}L_{\odot}^{pres},
\label{lumsolar2}
\end{equation}

where $L_{\odot}^{pres}$ is the luminosity of the Sun at present time
 $t_{\odot}$ ($\sim$ 4.6 Gyr).

For the future evolution of the solar luminosity, we have interpolated
the 
estimations given by 
Turck-Chieze \emph{et al}. (1988), which we fit with the relation:

\begin{equation}
L_{\odot}^{fut}(t)=[5.59\, Gyr\frac{1}{t}-1.39+0.26\, Gyr ^{-1}t]L_{\odot}^{pres}.
\label{lumsolfut}
\end{equation}

In Fig.   \ref{sunintime} we show the curve $L_{\odot}(t)$
 obtained for both cases.

\vspace{0.5cm}
\textbf{[Figure \ref{sunintime}]}

With Eq. \ref{lumsolar2} and \ref{lumsolfut}, in the next section we
calculate  the
evolution  of the habitable zones from 1 to 10 Gyr for several stars.

\section{The UV radiation fluxes around the extrasolar planets}\label{uvradiation}

By August 2005, there have been discovered 157 extrasolar planets in 138
planetary systems, 14 of which are multiple planetary systems (Schneider
2005). From these, only 23 stars (harbouring 32 different
planets) have been observed by IUE. We
have excluded HD150706 and HD99492 from our consideration because their
spectra are too
noisy.
The sample is listed in Table \ref{estpl}.

\textbf{[Table \ref{estpl}]}

The Hubble Space Telescope has also observed the UV spectra of Gliese
876,  HD114762 and HD209458, but the observed bandwidth (115-320 nm)
is  incomplete
to  include these
stars in our present analysis.  

We use  low dispersion (0.6 nm
resolution) and high dispersion ($\lambda/\Delta\lambda\sim$10000)
spectra, taken by the long wavelength cameras (LWP and LWR) in the
range  185-340 nm. The spectra are available from the IUE public
library (at     
 http:\emph{//ines.laeff.esa.es/cgi-ines/IUEdbsMY}), and have been
 calibrated  using the NEWSIPS (New Spectral Image
Processing System) algorithm. The
internal  accuracy of the high resolution calibration is around 4\%  
(Cassatella \emph{et al.} 2000) and  the errors of the low dispersion
spectra in the absolute calibration are around
10-15\% (Massa and Fitzpatrick 1998).

 In Fig.   \ref{hrturntar}, we
 plot  our sample of stars in an HR diagram, together with the
 criteria  used  by Turnbull and
 Tarter (2003a) to exclude stars that remain in the
 main sequence stage   less than 3 Gyr.
\vspace{0.2cm}

\textbf{[Figure \ref{hrturntar}]}

The giants HD219449 and HIP75458, and the subgiants HD27442 and
 $\gamma$  Cephei are above the curve M$_V$=-10[(B$-$V)-1.4]$^2$+6.5,
 one of the criteria used by Turnbull and Tarter.

Therefore, we exclude those stars from our analysis of habitability. 
 It is worth noting  that, although $\tau$ Boo,
$\rho$ Ind and 51 Peg are subgiants, their position in the HR
diagram puts them within the Turnbull-Tarter habitability criteria.


Different estimations of the ages of the stars of our sample are given
in Table \ref{estpl}. In the column 3, the
ages were estimated from the stellar luminosity (Lang 1992). From
Henry \emph{et al.} (1996) and Donahue (1993), we obtained the ages
associated  with
the chromospheric activity (column 4). In column 5 and 6, the values
were taken from Lachaume \emph{et al.} (1999) and N\"ordstrom \emph{et
  al.} (2004) respectively. 

It can be seen that there are large discrepancies between the
different estimated ages, that constitute an important source of
uncertainty in the determination of
the environmental stability of each exoplanet during their $\tau_{M
  life}$ and $\tau_{hab}$ lifetimes.

In Fig. \ref{hz_in_time_1}, we plot the evolution with time of the UV habitable
zone,  computed following the UV constrains found in Section
\ref{uvboundaries} together
with the traditional one, for each of the 17 stars of our sample and
for our solar system. For
the limits of the traditional HZ, we considered the intermediate
criteria, which considers the \emph{runaway greenhouse effect} for the
inner edge and the \emph{maximum greenhouse} for the outer one. For
the solar case, these limits are 0.84 AU and 1.67 AU, respectively. 
The visible flux was computed from the visible magnitude m$_v$ and the
stellar parallax listed in Table \ref{estpl}, and the limit was set as
the distance to the star where the radiation is the same as that at
0.84 AU and 1.67 AU of the Sun.

\textbf{[Figure \ref{hz_in_time_1}]}

 In the figure, each exoplanet is represented by a triangular
dot, indicating its orbital radius and  the present stellar  age.

\section {Discussion}\label{discussion}

The traditional HZs plotted in Fig.   \ref{hz_in_time_1} agree very
well with those obtained by Underwood \emph{et al.} (2003), who
computed  the
evolution in time of luminosity and effective temperature for several
main sequence stars, with masses between 0.5 and 1.5 M$_\odot$.  This
similarity supports our assumption that the stellar luminosity follows
a pattern similar to the one predicted by the Solar Standard Model.

In Fig.   \ref{hz_in_time_1} we see that, in most cases, the UV-HZ is
placed  much
closer to the central star than the traditional one. In those cases,
UV radiation inside the traditional HZ would not be an efficient source
for  photolysis, and
therefore the formation of the macromolecules needed for life would be
much more difficult, if not completely impossible.

To test the validity of our assumptions, in Fig. \ref{hz_in_time_1},
we applied our definition of the UV habitable zone to our solar system. The
results are consistent with previous analysis performed by Cockell (1998,
2000), Cockell \emph{et al.} (2000) and 
 C\'ordoba-Jabonero \emph{et al.} (2003).
For example, if the conditions of the atmospheric attenuation of ancient
Venus were \emph{similar} to Archean Earth, then the proximity to the
Sun determines that the UV flux must have been 1.9 times higher than the
Earth. The UV radiation environment would probably have been tolerable to
any potential life (Cockell 2000), as it is shown in Fig. \ref{hz_in_time_1}.
For the case of Mars the UV flux is 0.43 times the terrestrial one. The
lack of a significant ozone layer means that a significant portion of UV
spectrum reaches the surface of the planet without attenuation (see Table
\ref{uvbc} in the Appendix). Cockell \emph{et al.} (2000) showed that
the  present Martian
UV levels are about the same order of magnitude to the early terrestrial ones.

In near the 41\% stars of the sample: HD19994, 70 Vir, 14 Her,
55 Cnc,  47
UMa, $\epsilon$ Eri and  HD3651, there is no coincidence at
all  between the UV region and the HZ. 
In the cases of 51 Peg  and HD160691, the outer limit of the UV habitable
region  nearly coincides with the inner limit of the HZ. 
Of course, the limits plotted in Fig.  
\ref{hz_in_time_1} are not strict, and taking into account the error
in the  flux, both HZ and HZ$_{UV}$
could  coincide in a tiny region. However, considering the evolution
in time of both HZs, a planet orbiting in the habitable region would
remain  there
for less than 1 Gyr. We,
therefore,  include 51 Peg and HD160691 in the set of stars
where there is no overlap between the  UV region and the HZ. Something
similar happens to HD 186427 (16CygB).
This means that there is no coincidence between the HZ$_{UV}$ and the
HZ in almost  59 \% of the sample .

There are seven cases where the traditional  and the UV HZs overlap at
least partially and therefore can, in principle, host an habitable
planet in it. However in three cases (H147513, HD143761 and HD217014) the location of the giant
planet would make the orbit of the terrestrial planet unstable,
according to Turnbull and Tarter (2003a) and  Menou and Tabachnik
(2003).
 
There are five extrasolar
planetary systems ($\upsilon$ And, HD19994, HD143761, HD147513 and
HD160691) that have giant planets detected inside the traditional HZ. 
Assuming that there is a rocky
moon around these planets, we can check in Fig. \ref{hz_in_time_1}
(see also Table \ref{uvbc} in the Appendix)
if the UV radiation levels would be adequate
for the origin and evolution of life.   $\upsilon$
 And c , HD160691c, HD147513b and HD19994b have low UV
levels ( $\lesssim$ 2 times the Earth value), and therefore in these
cases a  hypothetical moon
will not be hostile for DNA-type life systems. However, UV radiation in
HD147513b, HD19994b and HD160691c are too low to trigger the formation
of life. 
In contrast, the  UV radiation on HD143761b would have
destroyed any  possible life on a rocky moon around the planet,
unless the  atmosphere filters 99 \% of the UV radiation (see Table
\ref{uvbc}).  Finally,
a moon orbiting $\upsilon$ And c would be  habitable according to both
criteria.

The two F stars of our
 sample (HD114762 and $\tau$ Boo) also present an interesting case.
 Around both stars, there is a region in the HZ where
complex  life would be burnt by UV radiation. 
In both cases, an atmospheric protection much larger than that of early Earth
would be needed to make the traditional HZ suitable for life. In
general, the   HZs around F
stars  are restricted by the damaging effect of UV radiation.

\section{Conclusions}\label{conclusions}

In this work we present a more restrictive criteria to habitability of
an extrasolar Earth-like planet
than the traditional liquid water one presented by Kasting \emph{et
  al.} (1993), as we analyze the  biological conditions to the origin and
the development of
life once the liquid-water scenery is already satisfied.

 Until an atmospheric protection would be built, a
planetary surface would be exposed to larger amounts of UV radiation,
which could act as one of the main source in the synthesis of
bioproducts and, in a certain wavelength, could be damaging for
DNA. Both  concepts represent the UV boundaries for the UV-HZ. 
In this work, we also analyze the evolution of UV habitability during
the main sequence life of the  star. 
We recognize that an atmospheric correction should be present as star and
a possible terrestrial planet age, but this point is beyond our scope.
However, the mediocrity factor contemplates an attenuation from either
the sea, the atmosphere, a rock, planetary obliquity, planetary rotation cycle, etc. of half
the radiation. 
Applying all these criteria to those stellar systems whose central star has
been observed by IUE, we obtained that an Earth-like planet orbiting
the stars   HD216437,
HD114752,
HD89744, $\tau$ Boo and Rho CrB could be habitable for at least 3 Gyr. 
A moon orbiting $\upsilon$ And c would be also suitable for life.
While, in the 59\% of the sample (51 Peg, 16CygB, HD160691, HD
,HD19994, 70 Vir, 14 Her, 55 Cnc,  47 UMa, $\epsilon$ Eri and  HD3651), the traditional HZ would not be
habitable following the UV criteria exposed in this work.

\acknowledgments
\textbf{Acknowledgements}

This research was supported by U401 and X271 UBACYT Research Projects from
the University of Buenos Aires and Project 03-12187 of the Agencia de
Promoci\'on Cient\'\i fica y Tecnol\'ogica (Argentina). A.B. was sponsored by a
CONICET graduate scholarship.
We thank Siegfried Franck, Jill Tarter, Margaret Turnbull and the  two
anonymous referees for their
useful comments and observations.
\appendix
\section{Appendix}
The UV radiation is usually divided in three wavelength ranges,
 UVA (315-400  nm), UVB (280-315 nm) and
UVC (190-280nm). In the Archean Earth, UVB and UVC would probably
have  reached the surface of the
Earth without attenuation (Sagan 1973, Kasting 1993). UV
 radiation have   played a major ecological role on the evolution and
 environmental distribution of cyanobacteria and their immediate 
 ancestors  during  evolutionary history. Garc\'\i a Pichel (1998) and
 Castenholz and Garc\'\i a Pichel (2000) showed that the biologically
 effective exposure to UVB, and not so much UVC, must have been
 particularly important.

Besides our previous analysis,
we may  also estimate the UVB and
UVC radiation values for the different extrasolar planets of our
sample at the semi-major orbital axis (Buccino \emph {et al.} 2004). 
In order to normalize the UVB and UVC radiation levels on the
extrasolar  planets, with the history of
life on our planet, the fluxes are expressed in Table \ref{uvbc} as
multipliers  of the Archean Sun UVB and UVC
outer space levels at 1AU, as well as N$^{*}_{DNA}$ values (without any
consideration  of the atmospheric or other natural attenuation, i.e. $\alpha$ $\sim$ 1).
In Table \ref{uvbc}, we present the UVB, UVC and N$^{*}_{DNA}$ fluxes
at the  semi-major axis (d$_{s-p}$) for a set
of extrasolar planets in which IUE observations were available.   
We have also included Venus,
Earth and Mars to compare our UV results within the Solar System
habitable zone.
Here the unit T.A.E.
represents the 
number of times the Archean
Solar radiation at 1 AU is available at each extrasolar planet.

The information provided in Table \ref{uvbc} is useful to compare the levels
of N$^{*}_{DNA}$ at different
planets with the values of UVB and UVC radiation that are the most
traditional
 way in which these
analysis are presented in the literature.

In the last table column we indicate if the planet is inside Kasting
\emph{et al.} (1993) habitable zone at present  (Y), or if it is outside
(N), or  if it will be inside it in the future (FY).   
If the central star is not suitable for life we leave an empty space.

Even though all these are giant planets, those with a (Y) sign, could
have  moons suitable for life as it was shown in the previous sections.

\textbf{[Table \ref{uvbc}]}

\newpage
\textbf{Figures}

\vspace{0.5cm}
\begin{figure}[htb!]
\centering
\includegraphics[width=0.7\textwidth]{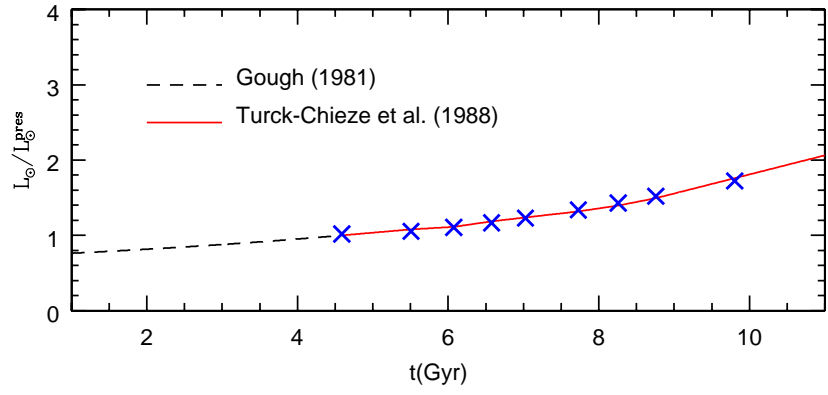}
\caption{Buccino \emph{et al.} Evolution of solar luminosity.}\label{sunintime}
\end{figure}
\vspace{0.5cm}
\newpage

\renewcommand{\baselinestretch}{1}
\begin{figure}[htb!]
\includegraphics[width=0.8\textwidth]{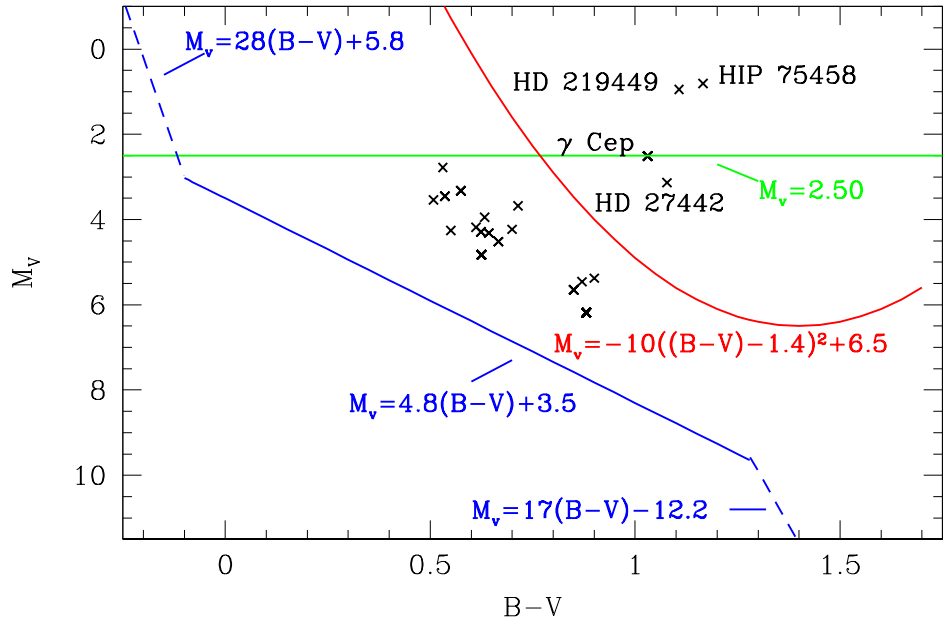}
\caption{Buccino \emph{et al.}. HR diagram.}\label{hrturntar} 
\end{figure}

\newpage
\renewcommand{\baselinestretch}{1}
\begin{figure}[htb!]
\hspace{-2cm}
\includegraphics[width=1.1\textwidth,height=0.9\textheight]{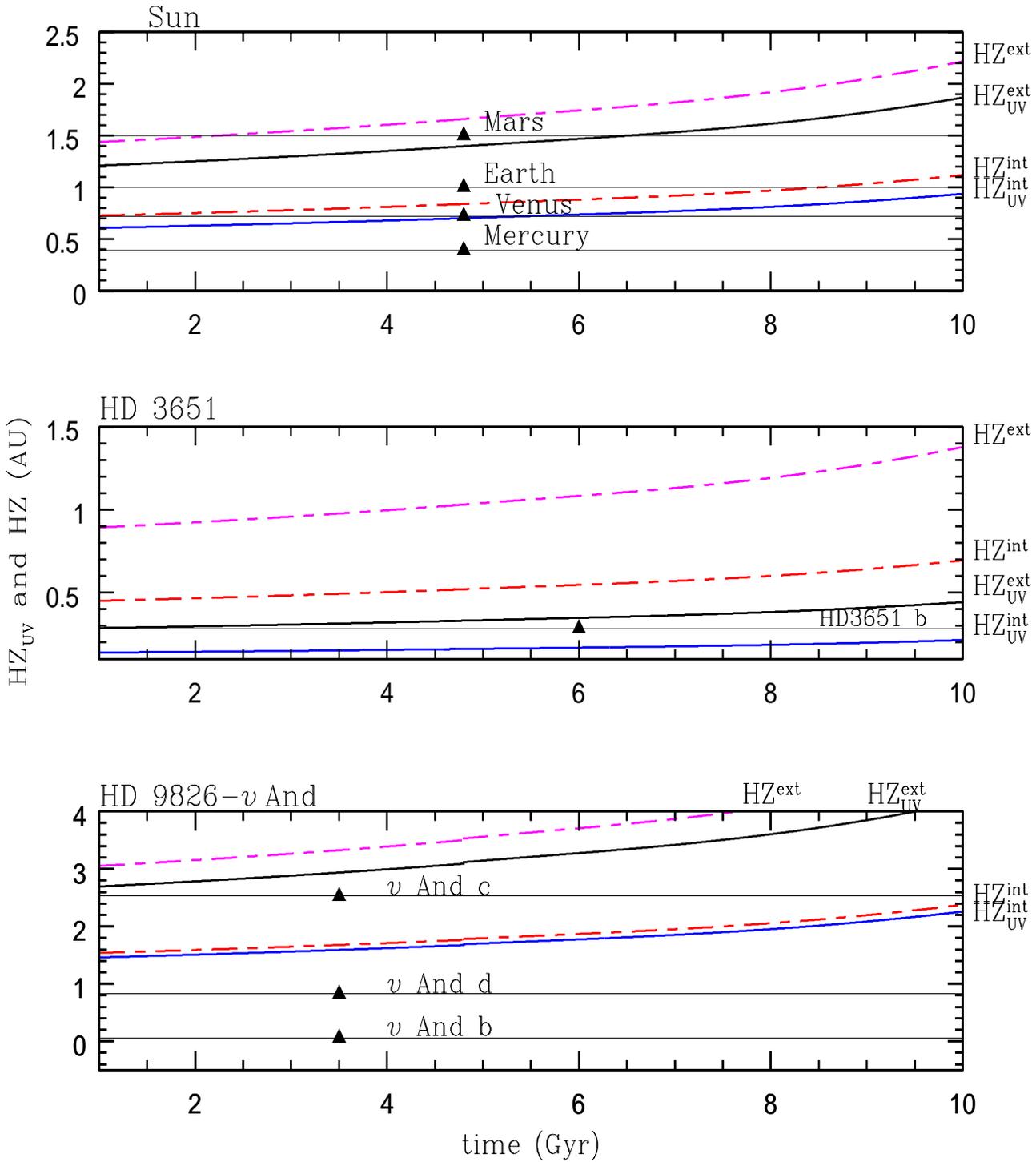}
\caption{ Buccino \emph{et al.}. HZ and UV-HZ.}\label{hz_in_time_1} 
\end{figure}

\renewcommand{\thefigure}{3}
\begin{figure}
\includegraphics[width=1.1\textwidth,height=0.9\textheight]{fig_2.epsi}
\caption{Cont. Buccino \emph{et al.}. HZ and UV-HZ.}
\end{figure}

\renewcommand{\thefigure}{3}
\begin{figure}
\includegraphics[width=1.1\textwidth,height=0.9\textheight]{fig_3.epsi}
\caption{Cont. Buccino \emph{et al.}. HZ and UV-HZ. }
\end{figure}

\renewcommand{\baselinestretch}{1}
\renewcommand{\thefigure}{3}
\begin{figure}
\includegraphics[width=1.1\textwidth,height=0.9\textheight]{fig_4.epsi}
\caption{\small{Cont. Buccino \emph{et al.}. HZ and UV-HZ.}}
\end{figure}

\renewcommand{\thefigure}{3}
\begin{figure}

\includegraphics[width=1.1\textwidth,height=0.9\textheight]{fig_5.epsi}
\caption{\small{Cont.Buccino \emph{et al.}. HZ and UV-HZ.}}
\end{figure}

\renewcommand{\thefigure}{3}
\begin{figure}
\includegraphics[width=1.1\textwidth,height=0.9\textheight]{fig_6.epsi}
\caption{\small{Cont. Buccino \emph{et al.}. HZ and UV-HZ.}}
\end{figure}

\newpage
\textbf{Tables}

\renewcommand{\baselinestretch}{1}

\begin{table}[htb!]

\centering
\makebox[5cm]{\small{
\begin{tabular}{|l l r r r r r r l l  c|}
\hline
\hline
             & Spectral  & &&  & & && & & \\

Stars        & type  &m$_v$& Plx  & Age$^1$& Age$^2$ & Age$^3$& Age$^4$&&& Other\\
        &and class & &\scriptsize{(mas)} & \scriptsize{(Gyr)}  &
        \scriptsize{(Gyr)}&\scriptsize{(Gyr)}  &\scriptsize{(Gyr)}&
        HIP   &HR & Names\\
\hline
\hline
Main sequence stars& &&&& &&&&&\\
\hline
Archean Sun & G2V& -26.8& 8749    & 1.0 &  &  & &&&\\
Present Sun & G2V & -26.8& 8749  & 4.6 &  &  &&&&\\
HD   3651 & K0V &5.9& 90.0& 15.1& 5.9&  & 17.0 & 3093& 166 &\\
HD   9826 & F8V &4.1&74.3& 4.3 &6.3&2.88&3.3 & 7513& 458 &$\upsilon$ And \\ 
HD  19994 & F8V & 5.1&44.7 &3.9 & 3.5&  & 4.7 & 14954& 962&\\          
HD  22049 & K2V &3.7&310.8& & 0.9&  &  & 16537& 1084 &$\epsilon$ Eri \\
HD  75732 & G8V &5.9& 79.8&13.4   & 6.46  &  &  &  43587 && 55 CnC\\  
HD  89744 & F7V &5.7 &25.7&2.8 & 4.5& &2.2&  50786& 4067 &\\
HD  95128 & G1V &5.0&71.0& 7.1 &7.0&6.3  &8.7 & 53721 & 4277 &47 \textbf{UMa}\\
HD 114762 & F9V &7.3&24.7 & 7.1  & & &11.8 & 64426 &  &\\
HD 117176 & G5V &4.9 &55.2&4.8 & 5.4&7.6&7.4 & 65721 & 5072&70 Vir \\
HD 143761 & G0Va & 5.4 &57.4&6.6 &7.2 & &12.1 & 78459 &  5968 &$\rho$ CrB \\
HD 145675 & K0V &8.8 &7.1&12.2  & 6.9& & & 79248 & &14 Her  \\
HD 147513 & G5V &5.4&77.7& 9.8 & & &8.5 & 80337 & 6094&\\
HD 160691 & G3IV-V &5.2 &65.5&6.6 & &6.2 & & 86796 &  6585 &$\mu$ Ara \\
HD 186427 & G3V &6.2&46.7&8.4 &7.4 & &9.9 & 96901 & 7504 &16 Cyg B \\
\hline
Subgiants&& &&&&&&&&\\
\hline
HD  27442 & K2IVa&4.4&54.8& 2.8& & &  & 19921 & 1355 &$\epsilon$ Ret\\    
HD 120136 & F6IV &4.5 &64.1 &4.5 & 4.2&   &2.4 & 67275 & 5185&$\tau$ Boo \\
HD 216437 & G2.5IV&6.0&37.7 &5.6  & &   &8.7 & 113137 & 8701&$\rho$ Ind \\
HD 217014 &G2.5IVa&5.5&65.1 &8.0 &5.7 &5.1  &9.2 & 113357& 8729&51 Peg \\
HD 222404 &K1IV&3.2&72.5&2.0& &  & & 116727& 8974&$\gamma$ Cep \\
\hline
Giants  &&& &&&&&&&\\
\hline
HD 137759 & K2III&3.3&31.9& &  &  & & 75458& 5744 & \\
HD 219449 & K0III&4.2&21.9& & & &  & 114855& 8841 &91 Aqr\\          
\hline
\end{tabular}}}

\caption{Planetary stars with IUE observation.} \label{estpl} 
\end{table}

\renewcommand{\baselinestretch}{1.}
\begin{table}[hp!]
\centering
\makebox[5cm]{\small{
\begin{tabular}{|l c r r r r c|}
\hline
\hline
           & &     & & & &\\
Planets & Star sp  & $d_{s-p}$ & \textbf{UVB} & \textbf{UVC} &\textbf{N$^{*}_{DNA}$}&Planet\\
       &  type and      & (AU)             & \scriptsize{(T.A.E)} &
\scriptsize{(T.A.E.)}& \scriptsize{(T.A.E.)}& within HZ\\    
             & class   &    & &&  &(Kasting\\
             & & &     & &  & \emph{et al.} 1993)\\
\hline
\hline
\small{Main  sequence stars} &&&&&&\\
\hline

Present Earth & G2V & 1.00 & 0.5 & 0.8 &1.4& (Y) \\
Mercury & G2V   & 0.39 & 3.3 & 5.3 &9.2& (N)\\
Venus   & G2V   & 0.72 & 0.9 & 1.5 &2.7& (N)\\
Mars   & G2V    & 1.50 & 0.2 & 0.4  &0.6&(Y)\\
HD   3651b & K0V & 0.28 & 2.7 & 3.9&1.9&(N) \\
HD   9826b & F8V & 0.06 & 1801.6& 2591.3&1660.1&(N)\\           
HD   9826c & F8V & 2.53 & 1.0& 1.4&0.9& (Y)\\
HD   9826d & F8V & 0.83 & 9.1& 13.1&8.4&(N)\\
HD   75732b & G8V &0.04& 172.2 &247.7 &123.2&(N)\\
HD   75732c & G8V &0.11 &20.6& 29.6&14.6&(N)\\
HD   75732d & G8V & 0.24&4.3 &6.2&3.1&(N)\\
HD   75732e & G8 V &5.90 &7E-3& 1E-2&5E-2&(N)\\
HD  19994b & F8V & 1.30 & 2.1  & 3.0 &1.8&(Y)\\          
HD  22049b & K2V & 3.30 & 1E-02 & 1E-02 &9E-2&(N)\\
HD  22049c & K2V & 40.00& 8E-06  & 1E-04 &6E-5&(N)\\
HD  89744b & F7V &0.88 & 14.8  & 21.3&12.2&(N)\\           
HD  95128b & G1V &2.10  &  0.4& 0.6 &0.3&(FY)\\
HD  95128c & G1V &3.73  & 0.1 & 0.2 &9E-2&(N)\\
HD  114762b & F9V & 0.30&41.4& 59.6 &66.1&(N)\\
HD 117176b & G5V & 0.48  & 6.9 &9.9 &4.3&(N)\\
HD 143761b & G0Va &0.22  & 53.6& 77.1&46.9&(N)\\
HD 145675b& K0V& 2.80& 3E-2& 4E-2&2E-2&(N)\\
HD 147513b & G5V &1.26  & 0.8  & 1.2 &0.8&(Y)\\
HD 160691b &G3IV/V& 1.50& 0.5& 0.8&0.5&(N)\\
HD 160691c &G3IV/V& 2.30& 0.2& 0.3&0.2&(Y)\\
HD 160691d &G3IV/V&4.17& 0.1& 0.1&5E-2&(N)\\
HD 186427b  & F7V &1.67& 0.4& 0.6 &0.4&(N)\\

\hline
\small{Subgiants}&&&&&&\\
\hline
HD  27442b & K2IVa& 1.18&0.5 &0.8&0.3&\\    
HD 120136b & F6IV & 0.05  & 3182.5& 4577.6&3720.1&(N)\\
HD 216437b &G2.5IV& 2.70& 0.3& 0.5&0.3&(N)\\
HD 217014b & G2.5IVa &0.05  & 381.7& 549.1&279.5&(N)\\
HD 222404b & K1IV& 2.03&0.4 &0.6 &0.2&\\
\hline
\small{Giants} & & &&&&\\
\hline
HD 137759b & K2III &1.34  & 2.2 &3.2&0.6&\\
HD 219449b & K0III& 0.30 & 51.0 &73.4 &28&\\
\hline
\end{tabular}}}
\caption{{UVB and UVC fluxes.}}\label{uvbc} 
\end{table}

\newpage
\textbf{Figures captions}

Figure \ref{sunintime}: Evolution of solar luminosity along main
sequence stage (Turck-Chieze \emph{et al}. 1988, and Gough 1981).

Figure \ref{hrturntar}: HR diagram representation of the  IUE-observed
  planetary stars in  Table \ref{estpl}, the habitability restrictions defined
  by  Turnbull and Tarter (2003a) are also included.

Figure \ref{hz_in_time_1}: The dashed lines represent the HZ
boundaries defined by Kasting   \emph{et al.} (1993) and the solid lines limit the region
where UV is suitable for life, following the criteria in section 
\ref{uvboundaries}. The triangles dot indicate the planets discovered in
each system.

\newpage
\textbf{Table captions}

Table \ref{estpl}: Planetary stars with IUE observation. Stellar ages
    were  determined
    according to (1) stellar luminosity (Lang 1992),  
    or to (2) chromospheric activity (Henry \emph{et al.} 1996,
    Donahue 1993),  (3) taken from Lauchame (1999)
    or (4) from N\"ordstrom \emph{et al.} (2004)

Table \ref{uvbc}: UVB and UVC fluxes at the semi-major orbital axis
(d$_{s-p}$) of a set of extrasolar planets in which IUE observations
were available. T.A.E: Times the solar radiation at 1 AU for
the Archean Earth.

\vspace{1cm}
\newpage
\textbf{Footnote}

Footnote 1: The  biological productivity is the amount of
  biomass  that is produced by photosynthesis per unit of time and per
  unit  of continental area.
\end{document}